\documentstyle[11pt,epsfig,amssymb,amsmath]{article}
\title{\bf  Brane cosmology in teleparallel and $f(T)$ gravity }
\author{K. Atazadeh$^{1,2}$ \hspace{-3mm} \thanks{Email: atazadeh@azaruniv.edu }\;\;\,and   A. Eghbali$^{1}$ \hspace{-1mm}\thanks{Email: a.eghbali@azaruniv.edu}
\\ $^1${\small \it Department of Physics,  Azarbaijan Shahid Madani University, 53714-161, Tabriz, Iran}
\\$^2${\small \it Research Institute for Astronomy and Astrophysics of Maragha (RIAAM), Maragha 55134-441, Iran}}
\begin{document}
\maketitle
\begin{abstract}
We consider the cosmology of a brane-world scenario in the framework of teleparallel and $f(T)$ gravity in a way that matter is localized on the brane.
We show that the cosmology of such branes is different from the standard cosmology in teleparallelism. In particular, we obtain a class of new solutions with a constant five-dimensional radius and cosmologically evolving brane in the context of constant torsion $f(T)$ gravity.
\end{abstract}

\section{Introduction}
Einstein had the idea of unifying gravitation and electromagnetism in 1928 \cite{Aldrovandi}. This attempt was based
on the mathematical structure of teleparallelism, also referred to as distant
or absolute parallelism. In other words, the idea was the introduction
of a tetrad field $-$ a field of orthonormal bases on the tangent spaces at each
point of the four-dimensional space-time. The tetrad has 16 components
whereas the gravitational field, represented by the space-time metric, has only
10. The six additional degrees of freedom of the tetrad was then supposed
by Einstein to be related to the six components of the electromagnetic field \cite{Aldrovandi}.
This attempt of unification did not succeed, because the additional six
degrees of freedom of the tetrad are actually eliminated by the six-parameter
local Lorentz invariance of the theory. However, Einstein introduced
concepts that remain important to the present day. Teleparallelism can be considered by using the
Weitzenb\"{o}ck connection, which has torsion, rather than the curvature defined by
the Levi-Civita connection \cite{Weitzenb}. The teleparallel
Lagrangian density is described by the torsion scalar, i.e, $T$. Recently, the authors have extended the Lagrangian density of teleparallel gravity, the so--called $f(T)$ gravity, in which various gravitational and cosmological solutions of this model are studied in \cite%
{Ferraro:2006jd}-\cite{Iorio:2012cm}. This concept is similar to the idea of $f(R)$ gravity.
Some people have studied brane world scenario in the framework of extended theories of gravitation such as $f(R)$ gravity \cite{ata}. Thus, inspired by these theories we became interested in studying brane cosmology within the teleparallelism theory.

Moreover, the idea that our world might be a brane embedded in a higher-dimensional space-–time (the bulk) \cite{horava} has been in the mainstream
of cosmological investigations in the past few years \cite{lisa}. This approach differs from the usual Kaluza–-Klein idea in that the size
of the extra dimensions can be large. The concept of large extra dimensions is discussed phenomenologically in \cite{bin2}. An important
ingredient of the brane world scenario is that matter is confined to the brane and the only communication between the brane and
the bulk is through gravitational interaction or some other dilatonic matter. In general, the matter on the brane leads to a cosmological
evolution which is different from the usual evolution governed by the Friedmann equation, that is, in brane cosmology the Hubble
parameter on the brane is proportional to the square of energy density \cite{brane}. This proportionality is a result of the application
of the Israel matching condition which is basically a relation between the extrinsic curvature and the energy–-momentum tensor
representing matter fields on the brane.  Extra dimension models of teleparallel and $f(T)$ gravity have recently been studied in \cite{nozari}.
In the present work, we study the teleparallel and $f(T)$ gravity in five-dimensions.

The organization of the paper is as follows: in section 2 we briefly review the teleparallel and $f(T)$ gravity in five-dimensions and
write the full system of equations. In section 3 we consider the cosmological equations for teleparallel and $f(T)$ gravity by imposing a constant torsion condition on the solutions. In the section 3.1 we study brane equations by inserting tension on the brane. Finally, we study the solutions with a constant five-dimensional radius in section 4. Conclusions are drawn in the last section.

\section{Brane cosmology in teleparallel and $f(T)$ gravity}
We consider a curvature-free brane embedded in a five-dimensional space-time (the bulk). We assume that our brane is located at $y=0$.
In  teleparallel gravity the tetrad components $e_{A}(x^{\mu})$ are the fundamental structures of the theory, where an index $A$ runs over $0,1,2,3,4$ for the tangent space at each point $x^{\mu}$ of the manifold. The relationship between tetrad and the space-time metric is given by
 \begin{eqnarray}\label{1}
 g_{\mu\nu}=\eta_{_{AB}}e_{~~\mu}^{A}\;e_{~~\nu}^{B},
 \end{eqnarray}
where $\mu$ and $\nu$ are Lorentzian (coordinates) indices on the manifold and run over $0, ..., 4$, and $\eta_{_{AB}}={\rm diag}[-1,1,1,1,1].$

In teleparallel gravity one uses the curvature-less Weitzenb\"{o}ck connection $\overset{{\rm w}}{\Gamma}^{~\rho}_{~\nu\mu}\equiv e_{A}^{~~\rho}\partial_{\mu} e_{~~\nu}^{A}$ \cite{Weitzenb} thus according to this connection the torsion $T^{\rho}_{~\mu\nu}$ and contorsion $K^{\mu\nu}_{~~\rho}$ tensors are, respectively, given by
\begin{eqnarray}\label{2}
 T^{\rho}_{~\mu\nu}&=&  \overset{{\rm w}}{\Gamma}^{~\rho}_{~\nu\mu}  - \overset{{\rm w}}{\Gamma}^{~\rho}_{~\mu \nu}= e_{A}^{~~\rho}(\partial_{\mu} e_{~~\nu}^{A}-\partial_{\nu} e_{~~\mu}^{A}),\\
 K^{\mu\nu}_{~~\rho}&=&-\frac{1}{2}(T^{\mu\nu}_{~~\rho}-T^{\nu\mu}_{~~\rho}-T_{\rho}^{~\mu\nu}).\label{3}
\end{eqnarray}
By using the above equations one can define the torsion scalar $T$ as follows
\begin{eqnarray}\label{4}
T= S^{~\mu\nu}_{\rho}T^{\rho}_{~\mu\nu},
\end{eqnarray}
in which
\begin{equation}\label{5}
S^{~~\mu\nu}_{\rho}=\frac{1}{2}( K^{\mu\nu}_{~~\rho}+\delta_{\rho}^{\;\mu}\;T^{\alpha\nu}_{~~\alpha}-\delta_{\rho}^{\;\nu}\;T^{\alpha\mu}_{~~\alpha}).
\end{equation}
In the context of $5$D $f(T)$ gravity,  we can write the Lagrangian in terms of torsion scalar, as \cite{Ben09}
\begin{eqnarray}\label{6}
I=\frac{1}{2\kappa^{2}_{5} } \int d^{^{5}}x \;e~ f(T),
\end{eqnarray}
where $e=|e|=$det$(e^{A}_{~~\mu})=\sqrt{-\det{g_{\mu\nu}}}$ and $\kappa^{2}_{5}=8\pi G_{5}$.\,\footnote{We have set units $8\pi G_{5}=1$.} In teleparallel gravity all gravitational fields are considered in the torsion tensor $T^{\rho}_{~\mu\nu}$, and torsion scalar, $T$, comes from it in a similar way as the curvature scalar, $R$, arises from the curvature (Riemann) tensor.

Variation of the action (\ref{6}), with respect to tetrad, gives the equations of motion \cite{Ben09}
\begin{eqnarray}\label{7}
\;f_{_{T}}[\partial_{\mu} (e\; e_{A}^{~\rho}\; S_{\rho}^{~\nu\mu})-e\; e_{_{A}}^{~\lambda}\;
S^{\rho\mu\nu} T_{\rho\mu\lambda}\; ]+f_{_{TT}} ~ e\; e_{_{A}}^{~\lambda}\; S_{\lambda}^{~\nu \mu} \partial_{\mu}T+ \frac{1}{2}  e~e_{_{A}}^{~\nu}\;f(T)~=~\Theta_{_{A}}^{\;\;\nu},
\end{eqnarray}
where $f_{_{T}} = \partial f(T) / \partial T$, $f_{_{TT}} = \partial^2 f(T) / \partial T^2$ and
$\Theta_{_{A}}^{\;\;\nu}$ is the energy-momentum tensor of the perfect fluid.

On the other hand, from the relation between the
Weitzenb\"{o}ck connection and the Levi-Civita connection
given by equation (\ref{3}), one can write the Riemann tensor for
the Levi-Civita connection in the form
\begin{eqnarray}\label{tensorR}
R^\rho_{\;\;\mu\lambda\nu}\!\!\!\!\!\!&&=\partial_{\lambda}\Gamma{}_{\;\mu\nu}^{\rho}
-\partial_{\nu}\Gamma {}_{\;\mu\lambda}^{\rho}
+\Gamma {}_{\;\sigma\lambda}^{\rho}\Gamma {}_{\;\mu\nu}^{\sigma}
-\Gamma {}_{\;\sigma\nu}^{\rho}\Gamma {}_{\;\mu\lambda}^{\sigma} \nonumber\\
&&=\nabla_\nu K^\rho_{\;\;\mu\lambda}-\nabla_\lambda K^\rho_{\;\;\mu\nu}
+K^\rho_{\;\;\sigma\nu}K^\sigma_{\;\;\mu\lambda}-K^\rho_{\;\;\sigma\lambda}K^\sigma_{\;\;\mu\nu}\;,
\end{eqnarray}
whose associated Ricci tensor can then be written as
\begin{equation}\label{eq8}
R_{\mu\nu}=\nabla_\nu K^\rho_{\;\;\mu\rho}-\nabla_\rho K^\rho_{\;\;\mu\nu}
+K^\rho_{\;\;\sigma\nu}K^\sigma_{\;\;\mu\rho}
-K^\rho_{\;\;\sigma\rho}K^\sigma_{\;\;\mu\nu}\;.
\end{equation}
Now, using $K^\rho_{\;\;\mu\nu}$ given by equation~(\ref{3}) along with the
relations $K^{(\mu\nu)\sigma}=T^{\mu(\nu\sigma)}=S^{\mu(\nu\sigma)}=0$ and considering that $S^\mu_{\;\;\rho\mu}=
2K^\mu_{\;\;\;\rho\mu}=-2T^\mu_{\;\;\;\rho\mu}$ one can get
~\cite{barrow}-\cite{reb}
\begin{eqnarray}\label{eq9}
R_{\mu\nu} &=& -\nabla^\rho S_{\nu\rho\mu}-g_{\mu\nu}\nabla^\rho T^\sigma_{\;\;\;\rho\sigma}
-S^{\rho\sigma}_{\;\;\;\;\;\mu}K_{\sigma\rho\nu}\;, \nonumber \\
R &=&-T-2\nabla^\mu T^\nu_{\;\;\;\mu\nu}\;,
\end{eqnarray}
and thus can obtain
\begin{equation}\label{eqdivs}
G_{\mu\nu}-\frac{1}{2}\,g_{\mu\nu}\,T
=-\nabla^\rho S_{\nu\rho\mu}-S^{\sigma\rho}_{\;\;\;\;\mu}K_{\rho\sigma\nu}\;,
\end{equation}
where $G_{\mu\nu}=R_{\mu\nu}-(1/2)\,g_{\mu\nu}\,R$ is the Einstein tensor.

Finally, by using equation~(\ref{eqdivs}), the field equations for
$f(T)$ gravity equation~(\ref{7}) can be rewritten in the form \cite{barrow}
\begin{equation}\label{eq11}
G_{\mu\nu}+\frac{1}{2f_{T}}\left(f(T)-Tf_{T}\right)g_{\mu\nu}+B_{\mu\nu}\frac{f_{TT}(T)}{f_{T}}=\frac{1}{f_{T}}\Theta_{\mu\nu},
\end{equation}
where we have defined $B_{\mu\nu}=S_{\nu\mu}\,^{\sigma}\,\nabla_{\sigma}T$.  When
$f(T)=T$, general relativity is
recovered, which verifies the claim that teleparallel gravity and general relativity are equivalent. In this case the field equations
are clearly covariant and the theory is also local Lorentz
invariant. In the more general case with $f(T)\neq T$ however, this is not the case.

For considering the cosmology of the model, we take the five-dimensional metric as follows\footnote{Note that the spatial curvature of the three-dimensional metric, $k$, is considered to be zero (spatially flat).}
\begin{eqnarray}\label{8}
ds^{2}=-A^{2}(t,y)dt^{2}+B^{2}(t,y)d\textbf{x}^{2}+C^{2}(t,y)dy^{2},
\end{eqnarray}
where $d\textbf{x}^{2}=dx_{1}^{2}+dx_{2}^{2}+dx_{3}^{2}$. With regard to the above relation, $g_{\mu\nu}={\rm diag}[-A^{2}(t,y), B^{2}(t,y),\\ B^{2}(t,y), B^{2}(t,y), C^{2}(t,y)]$ and then using equation \eqref{1}, the diagonal tetrad components read as
\begin{eqnarray}\label{9}
e^{_{A}}_{\;\mu}={\rm diag}[A(t,y), B(t,y), B(t,y),B (t,y), C(t,y)].
\end{eqnarray}
Now, by substituting equation \eqref{9} into the relation \eqref{2} and with the help of equations \eqref{3}-\eqref{5}, one can obtain
the value of the torsion scalar $T$ as follows\footnote{With regard to the mentioned notation in \cite{nour},  $\frac{\partial A}{\partial y}=A' \frac{d |y|}{d y}=A' (2\theta(y)-1), (\frac{\partial A}{\partial y})(\frac{\partial B}{\partial y})=A' B'$ and $\frac{\partial^2 A}{\partial y^2}=A'' +2A' \delta(y),$ where $\theta(y)$ is the Heaviside function;  $A'$ and $A''$ denote, respectively, the first and the second derivative of $A$ with respect to $|y|$. Note that $A''$ is the non-distributional part of the double derivative of $A$ (the standard
derivative) in which it vanishes on the brane. }:
\begin{eqnarray}\label{10}
T\;=\;\frac{6}{A^2 B^2 C^2}\left[-(AA')(BB')+ C^2 \dot{B}^2-
A^2 { B'}^2+(B\dot{B})(C\dot{C})\right],
\end{eqnarray}
where a dot  denotes  derivative with respect to $t$. In order to realize the $Z_{2}$ symmetry, the coefficients $A(t,y)$, $B(t,y)$ and $C(t,y)$ depend on $y$ through its modulus $|y|$. To survey the cosmological setup, we take the ${\Theta}^{\;\;\nu}_{\mu}$ as follows
\begin{eqnarray}\label{11}
{\Theta}^{\;\;\nu}_{\mu}=\frac{1}{C(y,t)}{\rm diag}[-\rho_b, p_b, p_b, p_b, 0]\delta(y),
\end{eqnarray}
where $\rho_b:=\rho_b(t)$ is the brane energy density and $p_b:=p_b(t)$ the brane pressure.  With this choice, the matter is localized on the brane. Furthermore, we require the equation of state of the matter on the brane to have the following form
\begin{eqnarray}\label{12}
p_b~=~ \omega \rho_b,
\end{eqnarray}
where $\omega$ is a real constant.

To continue, we study $f(T)$ gravity in the five-dimensions in the context of constant torsion regime. In this way, the equation of motion (\ref{eq11}) reduces to
 \begin{equation}\label{eq111}
{\cal{E}}_{\mu\nu}\;\equiv G_{\mu\nu}+\frac{1}{2f_{T}}\left(f(T)-Tf_{T}\right)g_{\mu\nu}=\frac{1}{f_{T}}\Theta_{\mu\nu}.
\end{equation}
Thus, the non-vanishing components of the equation of motion are ${\cal{E}}_{_{00}}, {\cal{E}}_{_{11}}={\cal{E}}_{_{22}}={\cal{E}}_{_{33}}$, ${\cal{E}}_{_{44}}$ and  ${\cal{E}}_{_{0 4}}$ that are, respectively, given by
\begin{eqnarray}
\frac{1}{C~f_{_{T}}} \rho_b \delta(y)&=&3 \Big[- {{1}\over {BC^2}}\Big(B''+2B'\delta(y) \Big)+{{B'}\over { BC^2}}\left({{C'}\over C} - {B' \over B}\right)+{\dot{B} \over {BA^2}}\left( {\dot{B} \over B}+{\dot{C} \over C}  \right) \Big]\nonumber\\
&&+\frac{1}{2} \Big(T-\frac{f(T)}{f_{_{T}}}\Big),\label{13-16}\\
\frac{1}{C~f_{_{T}}} p_b \delta(y) &=&  \Big[{{2}\over {BC^2}}\Big(B''+2B'\delta(y) \Big)+{{1}\over { AC^2}}\Big(A''+2A'\delta(y)\Big)+{{B'}\over B C^2}\left({{B'}\over B}- {2C' \over C}\right)\nonumber \\
 &&+{{A'} \over {AC^2}}\left({2{B'} \over B}-  {{C'} \over C} \right) -{\dot{B} \over {BA^2}}\left( {\dot{B} \over B}+{2\dot{C} \over C} \right)
+{\dot{A} \over {A^3}}\left( {\dot{C} \over C}+{2\dot{B} \over B} \right)\nonumber \\
 &&-{\ddot{C} \over CA^2}-{2\ddot{B} \over BA^2}\Big]-\frac{1}{2} \Big(T-\frac{f(T)}{f_{_{T}}}\Big),\label{13-16.6}\\
0&=&3\Big[{{B'}\over { B}}\left({{B'}\over B} + {A' \over A}\right)+ {{C^2 \dot{B}}\over { BA^2}}\left({\dot{A}\over A} - {\dot{B}\over B}\right)-
{{C^2 \ddot{B}}\over { BA^2}}\Big]-\frac{1}{2} {C^2} \Big(T-\frac{f(T)}{f_{_{T}}}\Big), \\
0&=&3\left( {\dot{B}A' \over {BA}}+  {{B'}\dot{C} \over { BC}}-{{\dot{B}}^{'} \over { B }}\right) \Big(2\theta(y)-1 \Big).\label{13-16.1}
\end{eqnarray}
With regard to matching the delta function on both sides of the first two equations ((\ref{13-16}) and (\ref{13-16.6})), we then obtain
\begin{eqnarray}\label{17}
\frac{{B'}_0}{B_0} \;=\; -\frac{1}{6 f_{_{T}}}C_0\rho_b,~~~~~~~~~~~~~~~~\frac{{A'}_0}{ A_0 }~=~\frac{1}{6 f_{_{T}}}C_0 \Big(2 \rho_b+3p_b\Big),
\end{eqnarray}
where $A_0:=A(t,0)$, $B_0:=B(t,0)$ and $C_0:=C(t,0)$. Once this matching is carried out, the delta function contributions cancel out
and the equations become valid everywhere. Also notice  that the obtained equation of state
is not of the form $p_b =\omega \rho_b$ but a time-dependent one.

\section{Teleparallel and $f(T)$ brane equations}

To obtain the Friedmann-like equation \cite{bin2} in constant torsion  $f(T)$  gravity, we first introduce the function \cite{nour}
\begin{eqnarray}\label{17.1}
F(t , y) \;=\; \frac{(B B')^2}{C^2}-\frac{(B \dot{B})^2}{A^2},
\end{eqnarray}
then, by assuming that equation (\ref{13-16.1}) is satisfied, the components ${\cal{E}}_{_{00}}$ and  ${\cal{E}}_{_{44}}$ of the
equation of motion can be written  in the following form
\begin{eqnarray}\label{17.2}
F'(t , y) \;=\; \frac{B^3 B'}{3}\Big(T-\frac{f(T)}{f_{_{T}}}\Big),~~~~~\dot{F} (t , y) \;=\; \frac{B^3 {\dot B}}{3}\Big(T-\frac{f(T)}{f_{_{T}}}\Big).
\end{eqnarray}
One can integrate the above equations and deduce the first integral of motion as
\begin{eqnarray}\label{17.3}
\frac{(B B')^2}{C^2}-\frac{(B \dot{B})^2}{A^2} \;=\; \frac{1}{12} B^4 \Big(T-\frac{f(T)}{f_{_{T}}}\Big)+{\cal C},
\end{eqnarray}
where ${\cal C}$ is an integration constant. Thus, using equation (\ref{17.3}), the function $A$ is entirely determined in such a way that
it is, in terms of $B$, $C$ and their derivatives, given by \cite{nour}
\begin{eqnarray}\label{17.4}
{A^2} \;=\; {\dot B}^2 \Big[\frac{{B'}^2}{C^2} - \frac{1}{12}B^2 \Big(T-\frac{f(T)}{f_{_{T}}}\Big)-\frac{{\cal C}}{B^2}\Big]^{-1}.
\end{eqnarray}
Finally, by evaluating the above equation at $y=0$ together with the use of the matching conditions (\ref{17}),  imposing
the equation of state (\ref{12}), and  considering the temporary gauge $A_0=1$, the Friedmann-like equation \cite{bin2} in constant torsion  $f(T)$  gravity
is obtained to be of the form
\begin{eqnarray}\label{17.4}
{H^2} \;=\;\frac {{\dot B}_0^2}{B_0^2} ~=~  \frac{1}{36 f_{_{T}}^2} \rho_b^2- \frac{\Lambda}{12} -\frac{{\cal C}}{B_0^4},
\end{eqnarray}
where $\Lambda=\Big(T-\frac{f(T)}{f_{_{T}}}\Big)$. Next,  we must examine the conservation of matter on the brane in teleparallel gravity. Thus, by imposing the matching conditions  on the ${\cal{E}}_{_{04}}$  at $y=0$ and taking $p_b= \omega \rho_b$, one gets  the conservation equation as
\begin{eqnarray}\label{19}
\dot{\rho_b}+3(1+w) \rho_b H \;=\;0.
\end{eqnarray}
Having solved the above equation, we obtain the energy density as follows
\begin{eqnarray}\label{20}
{\rho}_b\;=\;\rho_0 {{B}_0}^{-3(1+w)},
\end{eqnarray}
where $\rho_0$ is an integration constant.
Using the Friedmann-like equation (\ref{17.4}) with ${\cal C}=0$ and equation (\ref{19})  we find
\begin{eqnarray}\label{17.5}
\dot{\rho_b}^2 \;=\;\frac {(1+\omega)^2}{4}\rho_b^2 \Big(\frac{\rho_b^2}{f_{_{T}}^2} -3\Lambda\Big).
\end{eqnarray}
Here, we study $f(T)$ gravity in the context of constant torsion regime. Therefore, we must impose a constant torsion condition in our study. In this manner  we take $T=T_0=const.$ Now, using equations (\ref{17.4}) and (\ref{19}) and then inserting the matching conditions (\ref{17}) into
the right-hand side of equation (\ref{10}) as a constant value $T_0$, one can get on the brane
\begin{eqnarray}\label{17.55}
\frac{\dot{C}_0}{C_0} \;=\;\frac {(1+\omega)\rho_b}{2\dot{\rho}_b} \Big(\frac {(2+3\omega)}{6f_{_{T}}}\rho_b^2- \frac{\Lambda}{2}-T_0\Big).
\end{eqnarray}
To elaborate on our study, here, we will consider the cases corresponding to  $\Lambda=0, ~\Lambda <0$ and $\Lambda >0$ on the brane. In this way, the scale factor on the brane and the  deceleration parameter are calculated for all cases.\\

$Case ~i)$~~ For the choice $\Lambda=0$, i.e., $f(T)=T$,
the energy density on the brane as a function of the cosmic time is
\begin{eqnarray}\label{17.6}
{\rho_b}\;=\;-\Big(\beta_0 \pm \frac{(1+w)}{2} t \Big)^{-1},
\end{eqnarray}
where $\beta_0$ is an integration constant. By putting the above equation  into the equation (\ref{20}) we find the scale factor on the brane as follows:
\begin{eqnarray}\label{22}
{B}_0(t)\;\sim \;\Big(\beta_0 \pm \frac{(1+w)}{2} t \Big)^{\frac{1}{3(1+w)}},
\end{eqnarray}
From the above equation we deduce
\begin{eqnarray}\label{222}
{B}_0(t)\;\sim \;{t}^{\frac{1}{3(1+w)}},
\end{eqnarray}
thus, the accelerated brane universe occurred once the equation of state satisfied $w<-\frac{2}{3}$; this is the equation of state for dark energy.

To probe our model in the cosmological background, we look at the behavior of the deceleration parameter on the brane. The deceleration parameter, $q$, is enumerated as
\begin{eqnarray}\label{23}
q\;=\;-\frac{{B}_0 {\ddot{B}}_0}{{{\dot{B}}^2}_0},
\end{eqnarray}
where $B_{0}$ is scale factor on the brane. By using equation (\ref{22}), the equation (\ref{23}) can be written as
\begin{eqnarray}\label{24}
q\;=\;{3w+2}.
\end{eqnarray}
In the accelerating universe, $q$ is negative. To perceive the deceleration parameter,  we plot $q$ as a function of $w$ in Figure 1. From the above discussion, it can be seen that the obtained results for $f(T)=T$ correspond to teleparallel equivalence to general relativity, and the field equations reduce to the Einstein equations.  In addition, in this case ($\Lambda = 0$) the solution of equation (\ref{17.55}) is found  to be of the form
\begin{eqnarray}\label{24.11}
C_0(t)\;=\;\nu_0~\Big(\beta_0\pm \frac{1+\omega}{2}t\Big)^{-\frac{(2+3\omega)}{3(1+\omega)}} e^{\pm T_0(\beta_0 \pm \frac{1+\omega}{4}t)t},
\end{eqnarray}
where $\nu_0$ is an integration constant.
\begin{figure}
\begin{center}
\epsfig{figure=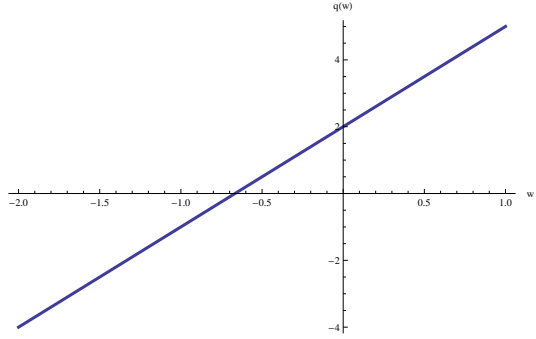,width=7cm}
\end{center}
\caption{\footnotesize The behavior of deceleration parameter $q$ as a function of $w$. An accelerating universe occurs for $w<-\frac{2}{3}$.}
\end{figure}\\

$Case ~ii)$~~ For  $\Lambda < 0$ we introduce  $\Lambda = - \eta^2$. In this case, the energy density on the brane from equation (\ref{17.5}) as a function of the cosmic time is given by
\begin{eqnarray}\label{17.7}
{\rho_b}\;=\;\pm  \frac{\sqrt{3} \eta f_{_{T}}}{\sinh  \Big(\frac{\sqrt{3} (1+\omega) \eta}{2} t \Big)}.
\end{eqnarray}
For the above solution, the scale factor on the brane takes the following form
\begin{eqnarray}\label{17.8}
B_0(t)\; \sim \; \Big(\frac{e^{\frac{\sqrt{3} (1+\omega) \eta} {2}t} - e^{-\frac{\sqrt{3} (1+\omega) \eta} {2}t}}{2}\Big)^{\frac{1}{3 (1+\omega)}}.
\end{eqnarray}
Inserting equation   (\ref{17.8}) into equation (\ref{23}), the deceleration parameter can be cast in the form
\begin{eqnarray}\label{17.9}
q\; = \; \frac{ 12(1+\omega) - \Big(2+e^{{\sqrt{3} (1+\omega) \eta} t} + e^{-{\sqrt{3} (1+\omega) \eta} t}\Big)}
 {\Big(2+e^{{\sqrt{3} (1+\omega) \eta} t} + e^{-{\sqrt{3} (1+\omega) \eta} t}\Big) }.
\end{eqnarray}
It can be seen that for $t\ll 1$, equation (\ref{24}) is recovered and this situation corresponds to  $\Lambda = 0$. For $t\gg 1$ we can get $q=-1$,
thus, in the late time we have the accelerating universe
(eternal de Sitter universe). Also, to calculate $C_0(t)$ one must substitute  equation (\ref{17.7}) into  (\ref{17.55}) together with $\Lambda = -\eta^2$.\\

$Case ~iii)$~~ For  the case corresponding to  $\Lambda > 0$ we put $\Lambda = \eta^2$.
Similar to the above, we obtain
\begin{eqnarray}\label{17.88}
B_0(t)\; \sim \; \sec^{-\frac{1}{3(1+\omega)}}\Big (\gamma_0 \pm  \frac{\sqrt{3} (1+\omega) \eta} {2}t \Big),
\end{eqnarray}
where $\gamma_0$ is a constant of integration. For this case,   the deceleration parameter is given by
\begin{eqnarray}\label{17.10}
q+1\; = \; {3(1+\omega)}~\csc^2\Big (\gamma_0 \pm  \frac{\sqrt{3} (1+\omega) \eta} {2}t \Big).
\end{eqnarray}
Again, one can use equations (\ref{20}), (\ref{17.55}) and (\ref{17.88}) to obtain $C_0(t)$ for this case.

Before proceeding to study  the brane equations with tension, we obtain a condition for accelerated expansion on the brane.
We show that in the limit $\rho_b^2\gg \frac{3}{2}  \Lambda  f_{_{T}}^2$, there is an accelerating universe.
To this end, we first write the component ${{\cal E}_{44}}=0$ of equation of motion at the position of the brane, $y=0$. Then, by using
the matching conditions (\ref{17}) and  the normalization $A_0=1$, we arrive at
\begin{eqnarray}\label{17.11}
\frac{ {\ddot{B}}_0}{{{\dot{B}}}_0}  + \frac {{\dot B}_0^2}{B_0^2} ~=~ - \frac{1}{36 f_{_{T}}^2} \rho_b \Big(\rho_b+3p_b\Big)- \frac{\Lambda}{6}.
\end{eqnarray}
Subtracting equation (\ref{17.4}) with ${\cal C} =0$ from equation (\ref{17.11}) gives
\begin{eqnarray}\label{17.12}
\frac{ {\ddot{B}}_0}{{{\dot{B}}}_0} ~=~ - \frac{1}{6f_{_{T}}^2} \Big[ \frac { 1}{3} \rho_b^2
+\frac {1} {2}  \rho_b p_b+ \frac{1}{2}  \Lambda f_{_{T}}^2 \Big].
\end{eqnarray}
Thus the condition for accelerated expansion on the brane is
\begin{eqnarray}\label{17.13}
{\ddot{B}}_0  >0,~~~~~if ~~~~p_b< -\Big( \frac{2\rho_b^2+ 3 \Lambda f_{_{T}}^2}{3\rho_b}\Big).
\end{eqnarray}
In the limit $\rho_b^2\gg \frac{3}{2}  \Lambda  f_{_{T}}^2$ we have an accelerating universe if $p_b <- \frac{2}{3} \rho_b$.

\subsection{The brane equations with tension}
In this subsection we shall consider a brane with total energy density $\rho_b = \rho_m + \lambda$, with $ \rho_m$ being the energy density
of the matter on the brane and $\lambda$ the constant tension of the brane. By considering  the cosmic matter
as a perfect fluid with equation of state $p_m = \omega \rho_m$ where $p_m = p_b + \lambda$, equation (\ref{19}) is expressed
\begin{eqnarray}\label{3.1.38}
3 {\dot{B}}_0  (\rho_m + p_m) + {{B}}_0 \dot{\rho_m}\;=\;0,
\end{eqnarray}
for which we have a solution similar to (\ref{20}). In the presence of $\rho_m$ and $\lambda$, equation (\ref{17.4}) with ${\cal C} = 0$ takes
the form
\begin{eqnarray}\label{3.1.39}
{H^2}  ~=~  \frac{1}{18 f_{_{T}}^2} (\frac{1}{2} \rho_m^2 + \rho_m  \lambda ) + \bar{\Lambda},
\end{eqnarray}
where $\bar{\Lambda} = \frac{\lambda^2}{36 f_{_{T}}^2} - \frac{\Lambda}{12}$. By introducing a new variable $x = B_0^{\bar{q}}$
in which $\bar{q}=3(1+\omega)$, equation (\ref{3.1.39}) is written as
\begin{eqnarray}\label{3.1.40}
\dot{x} ^2 ~=~ \bar{q}^2 \Big( \bar{\Lambda} x^2 + \frac{\lambda \bar{\rho_0}}{18 f_{_{T}}^2} x+ \frac{ \bar{\rho_0}^2}{36  f_{_{T}}^2}\Big).
\end{eqnarray}
where $\bar{\rho_0}$ is an integration constant in equation (\ref{3.1.38}). In order to complete the study of the cases corresponding to $\bar{\Lambda}  =0$, $\bar{\Lambda} >0$ and $\bar{\Lambda} <0$,
we explore the kinds of cosmology associated with the scale factor in (\ref{3.1.40}). Case $\bar{\Lambda}  =0$ means that there are some constant
$f(T)$'s which are satisfied in the following equation
\begin{eqnarray*}
T f_{_{T}}^2  - f(T)f_{_{T}}
= \frac{\lambda^2}{3}.
\end{eqnarray*}
In this case ($\bar{\Lambda}  =0$)  with the initial condition $B_0(0)=0$ for equation (\ref{3.1.40}) we have the following solution
\begin{eqnarray}\label{3.1.41}
B_0^{\bar{q}}(t) ~=~  \frac{\bar{q} \bar{\rho_0}}{6 f_{_{T}}} \Big(\frac{\bar{q} \lambda}{12  f_{_{T}}} t^2 +t\Big).
\end{eqnarray}
For cases $\bar{\Lambda} >0$ and $\bar{\Lambda} <0$, integration of equation (\ref{3.1.40}) with  $B_0(0)=0$ gives
\begin{eqnarray}\label{3.1.42}
B_0^{\bar{q}}(t) &=&  \frac{\bar{\rho_0}}{6 f_{_{T}}  \sqrt{\bar{\Lambda}}}~ \sinh  \Big(\bar{q}  \sqrt{\bar{\Lambda}} t\Big)+  \frac{\bar{\rho_0} \lambda}{36 f_{_{T}}^2  {\bar{\Lambda}}}\Big[\cosh \Big(\bar{q}  \sqrt{\bar{\Lambda}} t\Big)-1 \Big],~~~~~~~~~~~~~\bar{\Lambda} > 0,\\
B_0^{\bar{q}}(t) &=& \frac{\bar{\rho_0}}{6 f_{_{T}}  \sqrt{|\bar{\Lambda}}|} \sin \Big(\bar{q}  \sqrt{|\bar{\Lambda}|} t\Big)-   \frac{\bar{\rho_0} \lambda}{36 f_{_{T}}^2  {|\bar{\Lambda}|}}\Big[\cos \Big(\bar{q}  \sqrt{|\bar{\Lambda}|} t\Big)-1 \Big],~~~~~~~~~~\bar{\Lambda} < 0.
\end{eqnarray}
If we choose  $f(T) =T^n$, the constant torsion will be depended on $\lambda$. The results have been summarized in Table 1.

\vspace{5mm}
\begin{center}
{\scriptsize{ Table 1: }}\hspace{-2mm} {\scriptsize Constraints on $T$ and $n$ for cases $\bar{\Lambda} >0$ and $\bar{\Lambda} <0$ when $f(T) =T^n$.}\\
    \begin{tabular}{|l| l |l |    p{133mm} }
    \hline
   {\footnotesize$f(T) =T^n$ }& ~~{\footnotesize $T^{2n-1} < \frac{\lambda^2}{3 (n^2-n)}$ }& {\footnotesize
$~~T^{2n-1} > \frac{\lambda^2}{3 (n^2-n)}$}  \\ \hline
{\footnotesize $\bar{\Lambda} >0$} & ~~{\footnotesize $n \in (-\infty ~,~0) \cup (1~,~\infty)$}& {\footnotesize $~~0<n<1$} \\\hline
 {\footnotesize $\bar{\Lambda} <0$} & ~~{\footnotesize  $0<n<1$} &
{\footnotesize  $~~n \in (-\infty ~,~0) \cup (1~,~\infty)$}
\\ \hline
    \end{tabular}
\end{center}

\section{Solutions with a constant five-dimensional radius}
We consider the five-dimensional solution of the model by assuming the scale factor of the five-dimensional to be constant and normalized to  $ 1$ at all times.
We find a class of solutions with vanishing bulk matter and without a cosmological constant on the bulk.
In this respect, with $C(t, y)=1$,  ${{\cal E}_{04}}=0$ leads to
\begin{eqnarray}\label{255}
{A' \over {A}}\;=\;{{\dot{B}}^{'} \over { \dot{B} }}.
\end{eqnarray}
Integration gives
\begin{eqnarray}\label{25}
\dot{B}=Ag(t),
\end{eqnarray}
where $g(t)$ is an arbitrary function of $t$. Note that $g(t)=\dot{B}_{0}$ since $A_{0}=1$.
Furthermore, by inserting relations (\ref{255}) and \eqref{25} into the  component ${\cal{E}}_{_{00}}$ of the equation of motion, we find the following equation
\begin{eqnarray}\label{26}
(B^2)''- \frac{1}{3} \Lambda B^2  \;=\;2{g^2}(t).
\end{eqnarray}
As explained in section 3, we study $f(T)$ gravity in the context of constant torsion regime. Thus, to obtain the solutions of the model with $C(t,y)=1$
we must also impose  constant torsion condition on the solutions. By imposing that the torsion scalar is a constant and by considering equation (\ref{10})
with $C(t,y)=1$, we then get
\begin{eqnarray}\label{27.11}
T_{0}\;=\;6\left(-\frac{A'B'}{AB}+ \frac{\dot{B}^2}{A^2B^2}-
\frac{{ B'}^2}{{ B}^2}\right).
\end{eqnarray}
Then, by using the condition (\ref{17}) and regarding $H^2=\frac{\dot{B}_0^2}{B_0^2}$, with $H^2$ given by equation (\ref{17.4}) (with ${\cal C}=0$), we obtain the following equation
\begin{eqnarray}\label{27.11.1}
\frac {1}{6f_{_{T}}^2} (2+3\omega) \rho_b^2- \frac{\Lambda}{2}-T_0=0.
\end{eqnarray}
From the above equation $\rho_{_{b}}$ is found to be of the form
\begin{eqnarray}\label{27.11.2}
\rho_b^2~=~ \frac {(3T_0 f_{_{T}} - f(T))f_{_{T}}} {(\omega + \frac{2}{3})}.
\end{eqnarray}
We note that in this case $\rho_{_{b}}$ is constant and thus this equation restricts the solutions. As an example, for $\omega =-\frac{2}{3}$ one can obtain
$f(T) = T^{\frac{1}{3}}$.
Now, by solving  equation (\ref{26}) we obtain the solutions of the model for the cases corresponding to  $\Lambda=0, ~\Lambda >0$ and $\Lambda <0$.

$\bullet$ ~As mentioned in the preceding section, the case where $\Lambda =0$ corresponds to $f(T) = T$. For this case, integration with respect to $y$
and the use of equation (\ref{25}) gives
\begin{eqnarray}\label{27.1}
B^2 (t,y) ~=~  \zeta (t) |y| + g^2(t) y^2 + B_0^2,
\end{eqnarray}
where $\zeta (t)$ is an arbitrary function of $t$. One can use the first equation of (\ref{17}) with $C_0=1$ to determine the function
$\zeta (t)$. Thus,  utilizing equation \eqref{25} together with equation \eqref{19} leads to
\begin{eqnarray}\label{27}
B^2 (t,y) &=& B_0^2 \Big( 1- \frac{\rho_b}{3} |y| \Big) + g^2(t) y^2,\\
A (t,y) &=& \frac{B_0}{B}  \Big[ 1+ \frac{(\rho_b+ 3 p_b)}{3} |y| \Big] + \frac{\dot g(t)} {B}y^2.\label{27.1}
\end{eqnarray}
By substituting solutions (\ref{27}) and (\ref{27.1}) into
equation (\ref{27.11}) one can get $g^2(t)$ on the brane as follows:
\begin{eqnarray}\label{27.2}
 g^2(t)~=~\frac{B_0^2}{6} \Big(T_0 -\frac{(1+3\omega)}{3}\rho_b^2\Big),
\end{eqnarray}
where $\rho_b$ is given by the equation (\ref{27.11.2}).

$\bullet$ ~For the case  $\Lambda > 0$ we take $\Lambda=\eta^{2}$. Then, equation (\ref{26}) is written as

\begin{eqnarray}\label{28}
(B^2)''- \frac{1}{3} \eta^{2} B^2  \;=\;2{g^2}(t).
\end{eqnarray}
Thus, by solving equation (\ref{28}) one can get
\begin{eqnarray}\label{29}
B^{2}(y,t)=\psi(t)e^{\frac{\eta}{\sqrt{3}}|y|}+\phi(t)e^{\frac{-\eta}{\sqrt{3}}|y|}-\frac{6}{\eta^{2}}g^{2}(t),
\end{eqnarray}
where $\psi(t)$ and $\phi(t)$ are arbitrary functions of $t$. One can use the first equation of (\ref{17}) with $C_0=1$ to determine the functions
$\psi (t)$ and $\phi(t)$. Finally,  by using equations \eqref{25} and \eqref{19} we obtain \cite{hervik}
{\small \begin{eqnarray}\label{30}
B^{2}(y,t)&=&\left(B_0^{2}+\frac{6}{\eta^{2}}g^{2}(t)\right)\cosh\left(\frac{\eta|y|}{\sqrt{3}}\right)-\frac{B_0^{2}\rho_{_b}}{\sqrt{3}\eta f_T}\sinh\left(\frac{\eta|y|}{\sqrt{3}}\right)-\frac{6}{\eta^{2}}g^{2}(t),\\
A(y,t)&=&\frac{1}{B(y,t)}\left[\Big(B_0+6\dot{g}(t)/\eta^{2}\Big)\cosh(\eta|y|/\sqrt{3})+B_0\left(\frac{3p_{_b}+\rho_{_b}}{2\sqrt{3}\,\eta f_T}\right)\sinh(\eta|y|/\sqrt{3})-\frac{6}{\eta^{2}}\dot{g}(t)\right]\label{30.1},~~~
\end{eqnarray}}
where $\rho_b$ is given by the equation (\ref{27.11.2}). Also, to calculate $g(t)$ one must use  equations (\ref{30}), (\ref{30.1}) and  (\ref{27.11}) on the brane.
For the case $\Lambda=\eta^{2}>0$ when we choose $f(T)=T^{n}$, then, $T$ is positive if $n \in (-\infty ~,~0) \cup (1~,~\infty)$ and $T$ is negative if $0<n<1$.

$\bullet$ ~By taking $\Lambda=-\eta^{2}$ for case $\Lambda<0$, the general solution of equation (\ref{26}) is found to be of the form
\begin{eqnarray}\label{31}
B^{2}(y,t)=\bar{\psi}(t) \cos ({\frac{\eta |y|}{\sqrt{3}}})+\bar{\phi}(t) \sin ({\frac{\eta |y|}{\sqrt{3}}})+\frac{6}{\eta^{2}}g^{2}(t),
\end{eqnarray}
where $\bar{\psi}(t)$ and $\bar{\phi}(t)$ are arbitrary functions of $t$.  Similar to the  previous cases, we get
{\small \begin{eqnarray}\label{32}
B^{2}(y,t)&=&\left(B_0^{2}-\frac{6}{\eta^{2}}g^{2}(t)\right)\cos\left(\frac{\eta|y|}{\sqrt{3}}\right)-\frac{B_0^{2}\rho_{_b}}{\sqrt{3}\eta f_T}\sin\left(\frac{\eta|y|}{\sqrt{3}}\right)+\frac{6}{\eta^{2}}g^{2}(t),\\
A(y,t)&=&\frac{1}{B(y,t)}\left[\Big(B_0-6\dot{g}(t)/\eta^{2}\Big)\cos(\eta|y|/\sqrt{3})+B_0\left(\frac{3p_{_b}+\rho_{_b}}{2\sqrt{3}\,\eta f_T}\right)\sin(\eta|y|/\sqrt{3})+\frac{6}{\eta^{2}}\dot{g}(t)\right].
\end{eqnarray}}
Similarly, one can substitute the above equations into  (\ref{27.11}) to obtain $g(t)$ for this case.
\section{Conclusion}
In this paper, we have discussed $f(T)$ gravity in five dimensions within the context of a constant torsion regime. We have considered the cosmological equations for teleparallel and  $f(T)$ gravity  on the brane. We are aiming to further understand  the behavior of the brane universe studied  for the cases corresponding to  $\Lambda=0, ~\Lambda <0$ and $\Lambda >0$ on the brane.
In $case ~i)$ we have shown that  $f(T)=T$ corresponds to teleparallel equivalence to general relativity and the field equations reduce to the Einstein equations.
For $case ~i)$ ($\Lambda=0$)  the deceleration parameter as a function of $w$ has been plotted. From Figure 1, it can be seen that an accelerating universe occurs for $w<-\frac{2}{3}$. We have reached the result that, in the limit $t\ll1$, the case $\Lambda <0$ corresponds to  $\Lambda = 0$ and,  for $t\gg 1$, the declaration parameter approaches -1, namely  $q\rightarrow-1$.
Furthermore, we have  found a condition for accelerated expansion on the brane in the limit of $\rho_b^2\gg \frac{3}{2}  \Lambda  f_{_{T}}^2$, so an accelerating universe occurs  provided that $p_b <- \frac{2}{3} \rho_b$. Moreover, the scale factor on the brane in the presence of tension is obtained.
By imposing that the torsion scalar be a constant and by considering equation (\ref{10})
with $C(t,y)=1$, we have shown that for $\omega =-\frac{2}{3}$,  one possibility is
$f(T) = T^{\frac{1}{3}}$.
Finally, by imposing  constant torsion condition on the solutions we have obtained a class of solutions in the bulk in which the fifth dimension does not evolve dynamically.

\bigskip
\bigskip
\noindent {\bf Acknowledgments}

\bigskip
This work has been supported financially by Research
Institute for Astronomy and Astrophysics of Maragha
(RIAAM) under research project No.1/3720-53.\\
We would like to thank the anonymous referees for their invaluable comments and criticisms.

\end{document}